\documentclass{ws-p8-50x6-00}

\begin{document}

\title{Diquark condensation in two colour QCD}

\author{R. Aloisio and  A. Galante\footnote{Talk presented by A. Galante}}

\address{Dipartimento di Fisica dell'Universit\`a di L'Aquila,  
67100 L'Aquila, Italy\\E-mail: aloisio,galante@lngs.infn.it}

\author{V. Azcoiti}

\address{Departamento de F\'\i sica Te\'orica, Facultad 
de Ciencias, Universidad de Zaragoza, 50009 Zaragoza, 
Spain\\E-mail: azcoiti@azcoiti.unizar.es}

\author{G. Di Carlo}

\address{Istituto Nazionale di Fisica Nucleare, 
Laboratori Nazionali di Frascati, P.O.B. 13 -  00044 Frascati, 
Italy\\E-mail: gdicarlo@lnf.infn.it}

\author{A. F. Grillo}

\address{Istituto Nazionale di Fisica Nucleare, 
Laboratori Nazionali del Gran Sasso, 67010 Assergi (L'Aquila), 
Italy\\E-mail: grillo@lngs.infn.it}


\maketitle

\abstracts{
Unquenched lattice SU(2) is studied at nonzero chemical potential in the
strong coupling limit. The topic of diquark condensation is addressed
analyzing the probability distribution function of the diquark condensate.
We present results at zero external source without using any
potentially dangerous extrapolation procedure.
We find strong evidences for a (high density) second order phase
transition where a diquark condensate appears, and show
quantitative agreement of lattice calculations
with low-energy effective Lagrangian calculations.
}

\section{Introduction}

Recently the standard scenario for the phase diagram of QCD
in the chemical potential-temperature plane has changed;
besides the standard hadronic and quark-gluon plasma phases,
the existence of a new state of matter has been claimed by several 
groups\cite{wil}.
This new phase is characteristic of the high density-low temperature 
regime; the asymptotic freedom of QCD and the known instability
of large Fermi spheres in presence of (whatever weak) attractive forces
results in a pairing of quarks of higher momenta (in analogy with the 
Cooper pairing in solid state systems at low temperature).
A condensation of quark pairs should be the distinctive signal of the
new phase and has been indeed predicted using simplified 
phenomenological models of QCD.

Unfortunately the lattice approach, the most powerful tool to perform
first principles non perturbative studies, is affected in the case of 
finite density QCD by the well known sign problem that has prevented
until now any step towards the understanding of this new physics.
This is not the case for the SU(2) theory where the fermions are in the
pseudo-real representation and finite density numerical simulations 
are feasible.

In the following we present a detailed study of diquark condensation
for the unquenched two colours model. 
This work follows a previous one
where we considered chiral and diquark susceptibilities and found
strong evidence for a phase transition separating the ordinary
low density phase from a high density one where chiral symmetry
is restored and the baryon number symmetry is broken\cite{noi1}.
In the present work we mainly focus on an approach, based on 
the analysis of the probability distribution function (p.d.f.) of 
the order parameter, to extract the value of the order parameter directly 
at zero external source without using any potentially dangerous 
extrapolation procedure, implicit in the standard approach.

The study of diquark condensation in two colours QCD is an interesting 
topic by itself and, notwithstanding the relevant differences
between SU(2) and SU(3) at non zero baryon density, we can hope to use 
some of the results to get insights about the three colours case.
Even if the phase diagram of SU(3) and SU(2) are expected
to share similarities, we have to be aware of the differences 
between the two models.
\begin{description}
\item 
$\bullet$ 
The quark-quark condensate $\langle qq\rangle$ is coloured
for the $N_c=3$ theory and colourless for the $N_c=2$ one.
This implies that in the SU(3) case the diquark condensation has to be
interpreted as an Higgs-like mechanism leading to the breaking of the local 
colour symmetry, while in the SU(2) case we have a standard spontaneous
symmetry breaking (SSB) of the $U_V(1)$ symmetry associated with conservation
of baryon number.
\item 
$\bullet$ 
The zero temperature critical chemical potential, 
related to the mass of the lightest baryonic state, is different in the
two cases:
it is $1/3$ of the nucleon mass for SU(3) and $1/2$ of the pion mass
for SU(2) (i.e. zero at vanishing quark mass). 
For $N_c=2$, in order to have a phase diagram similar
to the SU(3) one, it is appropriate to consider a non zero bare quark mass.

\item
$\bullet$ 
The lightest baryonic state is a fermion for the tree colours
model and a boson for the two colours one.
\end{description}

Even if our results can not be directly extended to the SU(3) case
we can use lattice calculations to do a highly non trivial, quantitative
check of the continuum predictions for the SU(2) theory. In particular we
have considered results coming from low energy effective
Lagrangian calculations\cite{lel}.

In the next section we present some details of the algorithm. 
The last section is devoted to the presentation and discussion
of numerical results.

\section{Overview of the numerical technique}

\subsection{Simulation scheme}
The standard way to study SSB is to introduce first an explicit symmetry
breaking term in the action. If we do that for the diquark in the SU(2)
model we have to add a term\cite{h}
$j(\psi\tau_2\psi + \bar\psi\tau_2\bar\psi)$
and, after integrating the Grassmann field, the fermionic contribution
for $N_f=4$ quark flavours
becomes proportional to the Pfaffian of a $4V\times 4V$ matrix\cite{noi1}:
\begin{equation}\label{1}
{\cal Z}_{ferm}(j) = Pf(B+j) = \pm \sqrt{\det(B+j)}
\end{equation}
where $B$ is
\begin{equation}\label{2}
 B=
\left(\begin{array}{cc}
0 & \frac{1}{2}\Delta\tau_2 \\
-\frac{1}{2}\Delta^T\tau_2 & 0 \end{array} \right).
\end{equation}
and $\Delta$ is the usual lattice Dirac operator (it contains the mass
and $\mu$ dependence).
Using the relation $\tau_2\Delta\tau_2=\Delta^*$ we can easily prove
that $B$ is antihermitian and $\det(B+j)\ge 0$ for any $j$. 

It can also be shown that the eigenvalues of $B$ are doubly degenerate
and $B^2$ is in a block diagonal form with two hermitian blocks 
on the diagonal having the same eigenvalues. 
It follows that to compute $\det(B+j)$ for any $j$ (i.e. to
obtain all the eigenvalues of $B$)
it is sufficient to diagonalize only one block of $B^2$ 
(reducing the problem to the diagonalization of a $2V\times 2V$ 
hermitian matrix) and then take the two pure imaginary square 
roots of the (real and negative) eigenvalues.

To avoid the sign ambiguity in (\ref{1}) it is customary to consider 
a theory with $N_f=8$ quark flavours where the fermionic
partition function becomes ${\cal Z}_{ferm}(j)=\det(B+j)$.
On the other side if we are able to work directly at zero
diquark source things become much simpler.
In the $j=0$ limit the sign ambiguity disappears and the Pfaffian
is positive definite
since $Pf(B)\equiv \det\Delta$ and the last quantity is real
and positive for any value of $\mu$.
Then we can easily consider any value of $N_f$ writing  
${\cal Z}_{ferm}(j=0)=(\det(B))^{N_f/8}$. 
In the next section we show how it is possible to take advantage of this 
feature and extract the value of the order parameter for the 
diquark condensation for all $N_f$ without any extrapolation to $j=0$.

We have studied the phase
structure of the theory in the limit of infinite gauge coupling ($\beta=0$).
To simulate the $\beta=0$ limit of the theory we have measured fermionic
observables on gauge configurations generated randomly, i.e. with only the 
Haar measure of the gauge group as a weight.
This choice implies a Gaussian distribution of the plaquette energy
around zero which, according to the results of Morrison and Hands\cite{h}, 
has a net overlap with the importance sample of gauge configurations
at the values of $\mu$ and $m$ used in our calculations.
The validity of this procedure has also been tested comparing
different physical observables (number density and chiral condensate)
with Hybrid Montecarlo results\cite{mdp}.

We have considered the theory in a $4^4$ and $6^4$ lattice diagonalizing $300$ 
gauge configurations in the first lattice volume and $100$ in the second one.
As we pointed out in the introduction, to stay closer to the SU(3) 
case the simulations have been performed at non zero quark mass.
We choosed $m=0.025,0.05,0.20$ and values of the chemical potential 
ranging from $\mu=0$ to $\mu=1.0$.

All numerical simulations have been performed on a cluster of Pentium II 
and PentiumPro at the INFN Gran Sasso National Laboratory.

\subsection{Analysis of the probability distribution function}

The use of the p.d.f. to analyze the spontaneous symmetry breaking
in spin system or Quantum Field Theories with bosonic degrees of
freedom is a standard procedure. Less standard is its application to
QFT with Grassmann fields where, for obvious reasons, the fermionic 
degrees of freedom have to be integrated analytically.
Nevertheless this method has been developed to extract the chiral
condensate in the chiral limit from simulations of QFT with
fermions\cite{pdc}.

The same ideas can be used to study the vacuum structure of two
colour QCD at non zero density and specifically to extract the
diquark condensate at $j=0$. We refer to the original paper
for a full description of the p.d.f. technique and  present
a brief introduction focusing on the peculiarities of the diquark
condensate case.

Let $\alpha$ be an index which characterizes all possible
(degenerate) vacuum states and $w_\alpha$ the probability to get the
vacuum state $\alpha$ when choosing randomly an equilibrium state.
If $c_\alpha$ is the order parameter in the $\alpha$ state we can
write
\begin{equation}\label{3}
c_\alpha=\langle\frac{1}{V}\sum_x \psi\tau_2\psi (x)+
\bar\psi\tau_2\bar\psi (x) \rangle_\alpha
\end{equation}
where $V$ is the lattice volume and the sum is over all lattice points.
$P(c)$, the p.d.f. of the diquark order parameter $c$, will be given by
\begin{eqnarray}
P(c) &=& \sum_\alpha w_\alpha\delta (c-c_\alpha)\nonumber\\
&=& \lim_{V\to\infty} \frac{1}{\cal Z} 
\int [dU][d\bar\psi] [d\psi]
e^{-S_G(U)+\bar\psi\Delta\psi}
\delta (\frac{1}{V}\sum_x \psi\tau_2\psi (x)+
\bar\psi\tau_2\bar\psi (x)-c)\nonumber
\end{eqnarray}

The main point is that while $P(c)$ is not directly accessible 
with a numerical simulation its Fourier transform 
\begin{equation}\label{4}
P(q) = \int dc e^{iqc} P(c)
\end{equation}
can be easily computed. Inserting in (\ref{4}) the definition of $P(c)$
and using an integral representation for the $\delta$-function, after 
some algebra we arrive at the following formula
\begin{equation}\label{5}
P_V(q)=\frac{1}{{\cal Z}(j=0)}
\int [dU] e^{-S_G(U)} \frac{Pf B(j=\frac{iq}{V})}{Pf B(j=0)}
[Pf B(j=0)]^{\frac{N_f}{4}}
\end{equation}
where $P_V(q)$ is the Fourier transformed p.d.f. of the diquark order
parameter at zero external source and finite volume.

To take advantage of (\ref{5}) we should be able to compute correctly
the Pfaffians involved. This indeed turns out to be easy.
First note that, as previously stated, $Pf B(j=0)$ is a real and positive
quantity so that the only ambiguity is related to the sign of 
$Pf B(j=\frac{iq}{V})=\pm\sqrt{\det B(j=iq/V)}$. If we remember that the
$4V$ eigenvalues of the antihermitian matrix $B$ are doubly degenerate and  
can be written as $\pm i\lambda_n$ ($n=1,\cdots,V$)
with real and positive $\lambda_n$ we arrive at the following expression
\begin{equation}\label{6}
Pf B(j=\frac{iq}{V}) = \pm \prod_{n=1}^{V} (\lambda_n^2-\frac{q^2}{V^2})
\end{equation}
The sign ambiguity can now be solved noticing that (\ref{6}) has to be
positive at $j=0$ and, increasing $q$, changes sign each time $j$
is equal to one of the eigenvalues. The only possible exception
would be an (accidental) extra degeneracy in the eigenvalues of $B$ 
resulting in the Pfaffian not crossing zero but tangent to the 
horizontal axis.
This situation never occurred in our simulations.

Once we have the $P_V(q)$ and hence, by simple Fourier transform,
the p.d.f. of the order parameter we need to extract the correct
value for the order parameter. To do that we first have to recall
that diquark condensate is the order parameter of the $U_V(1)$ 
symmetry associated with baryon number conservation. 
Indeed $\langle\psi\tau_2\psi+\bar\psi\tau_2\bar\psi\rangle$ 
and $i\langle\psi\tau_2\psi-\bar\psi\tau_2\bar\psi\rangle$ are the
components of a vector (in a plane) which rotates by
an angle $2\theta$ when we do a global phase transformation of parameter 
$\theta$ on the fermionic fields. Therefore, if $c_0$ is the 
vacuum expectation value of the diquark condensate corresponding
to the $\alpha$-vacuum selected when switching-on a diquark
source term, $P(c)$ can be computed as
\begin{equation}\label{7}
P(c) = \frac{1}{2\pi} \int d\alpha \delta(c-c_0\cos (2\alpha))
\end{equation}
which gives $P(c)=1/(\pi(c_0^2-c^2)^{1/2})$ for $-c_0\le c\le c_0$
and $P(c)=0$ otherwise\cite{pdc}. In the symmetric phase $c_0=0$ and $P(c)$
reduces correctly to a $\delta$-function in the origin.

\begin{figure}[t]
\epsfig{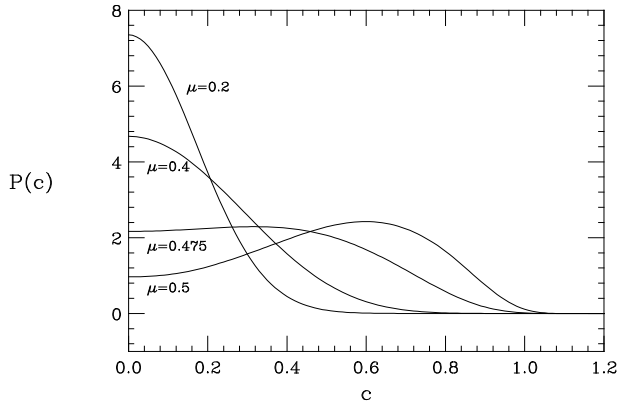} 
\caption{$P(c)$ for the $4^4$ lattice at $\mu=0.2, 0.4, 0.475, 0.5$
(from top to bottom).
\label{fig1}}
\end{figure}

The above results are valid in the thermodynamic limit while, at finite
volume, the non analyticities of the p.d.f. are absent.
Without entering in the details of a finite size scaling analysis 
we expect, for the finite volume p.d.f. $P_V(c)$, 
a function peaked in the origin in the symmetric phase and  peaked
at some non zero value in the broken phase. 
This is indeed the behaviour we can observe in fig. \ref{fig1} where the 
the p.d.f. of the smallest volume is reported at different values
of $\mu$. It is clear that, increasing the chemical potential,
the vacuum starts to be degenerate signalling a spontaneous breaking
of the baryon number conservation.

We can also compare the $P_V(c)$ for two lattice volumes in the symmetric
and broken phase. This is done in fig. 2 where we see clearly as,
increasing the volume, the peak of the p.d.f. becomes sharper.
To determine the value of the diquark condensate we used the position
of the peak: a definition that clearly converges to the correct value
in the thermodynamic limit.

\begin{figure}[htbp]
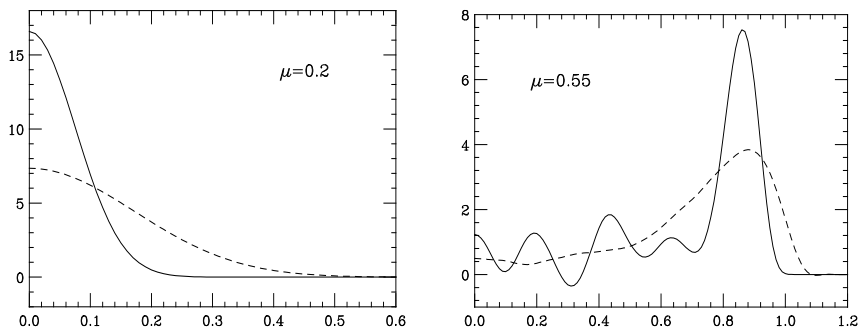

\centerline{\epsfysize=40mm 
\epsfig{file=fig2a.epsi,angle=90,width=150pt,height=120pt} 
\hspace*{5mm}
\epsfysize=40mm 
\epsfig{file=fig2b.epsi,angle=90,width=150pt,height=120pt} 
}
\vspace*{-3mm}
\caption{P.d.f. for $4^4$ (dashed line) and $6^4$ (continuous line) in the
symmetric (left) and broken (right) phase.
\label{fig2}}
\end{figure}

From fig. (\ref{fig2}) we see also that data of the larger 
volume are more noisy
(indeed we also get negative values for the p.d.f. in the broken phase).
To have an estimate of the errors we calculated several distribution 
functions for the $6^4$ lattice using independent subsets of our data.
We saw that the position of the peak was very stable, within 1 percent
in the broken phase, and we used this quantity for the errorbars.

\section{Results}

Here we present the results for the diquark condensate at $j=0$ as a function
of the chemical potential. Fig. 3 contains our data points for the
$6^4$ volume for several values of $N_f$ at two different masses:
$m=0.025$ and $m=0.2$ .

\begin{figure}[htbp]
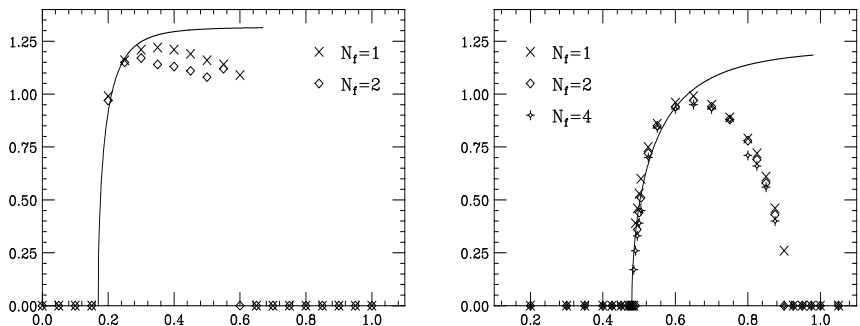

\centerline{\epsfysize=40mm 
\epsfig{file=fig3a.epsi,angle=90,width=150pt,height=120pt} 
\hspace*{5mm}
\epsfysize=40mm 
\epsfig{file=fig3b.epsi,angle=90,width=150pt,height=120pt} 
}
\vspace*{-3mm}
\caption{Diquark condensate as a function of $\mu$
for $6^4$ lattice (points) and the 
continuum predictions (line) for $m=0.025$ (left) and $m=0.2$
(right).
\label{fig3}}
\end{figure}

First we can derive a peculiar phase diagram: two symmetric phases
separated by a broken one and two possibly continuous transition points. 
This result is not new and has already
been predicted by Mean Field calculations\cite{mf} 
as well as numerical calculations\cite{noi1}.
The high density symmetric phase has no physical relevance, it is simply the 
consequence of the saturation of all lattice sites with quarks.
This saturation phenomena has nothing to do with continuum physics
and is a pure lattice artifact. 
The physically interesting phase transition, i.e. the transition that has a
continuum counterpart, is the first one.

To test our numerical results we have considered the analytical predictions
of low energy effective Lagrangian. The $j=0$ case is described by a simple 
formula where the only two parameters are the
chiral condensate and one half the pion mass calculated at $j=0, 
\mu=0$\cite{lel}.
We did independent simulations to calculate this quantities at the
same values of $m$ used in the previous analysis, 
obtaining the values in Table 1.
Another prediction of the continuum calculation is that the value of the
diquark condensate is independent of the flavour number.
\begin{table}[t]
\caption{Parameters for the low energy effective Lagrangian predictions.
\label{tab}}
\begin{center}
\footnotesize
\begin{tabular}{|c|c|c|}
\hline
$m$ & $\langle\bar\psi\psi\rangle$ & $m_\pi/2$\\
\hline
0.025 & 1.31(2). &0.1696(11)\\
\hline
0.2  &  1.23(2)  &  0.4841(7)\\
\hline
\end{tabular}
\end{center}
\end{table}

The continuous line in fig. 3 is the analytical prediction for
this model. We see a remarkable agreement with our strong coupling
results up to values of the chemical potential where the saturation
effects start to be relevant. We see also that the $N_f$ dependence
in the simulation points is small.

We checked explicitly the
low energy effective Lagrangian prediction at finite $j$ too.
In that case we used standard techniques to compute
chiral and diquark condensate as well as the number
density\cite{noi1} in a $N_f=8$ simulation (fig. 4).
Also in that case (where no phase transition is present
and the predictions are still independent of $N_f$ ) 
the agreement is very good up to values of $\mu$ where the 
number density becomes a consistent fraction of its maximum
lattice value (i.e. $\mu/m_\pi\simeq 0.6$).

\begin{figure}[t]
\epsfig{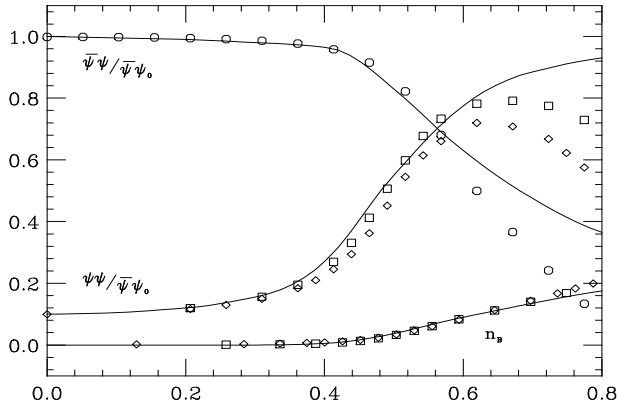} 
\caption{ Low energy Lagrangian predictions (lines) and our $N_f=8$
data (symbols)
for chiral condensate, diquark condensate and number density at $m=0.2$
and $j=0.1m$ vs. $\mu/m_\pi$. 
$6^4$ data are squares and circles, $4^4$ data are diamonds.
\label{fig4}}
\end{figure}

It is surprising that a $\beta=0$ calculation has a so good agreement
with a continuum prediction. Since the analytical predictions are,
at the values of $\mu$ presented, well inside the validity region of the 
low energy approximation ($\mu<<$ the mass of the first non goldstone
excitation) we can conclude that this gives an indication of a very
small dependence of lattice results on $\beta$.

\section*{Acknowledgments}
This work has been partially supported by CICYT (Proyecto AEN97-1680)
and by a INFN-CICYT collaboration. The Consorzio Ricerca
Gran Sasso has provided part of the computer resources needed for
this work.

\end{document}